\documentclass[9pt, sigconf, nonacm]{acmart}

\usepackage{algorithm}
\usepackage{balance}
\usepackage[noend]{algpseudocode}
\usepackage{caption}
\usepackage{subcaption}
\usepackage{xcolor}
\usepackage{varwidth}
\usepackage{mdwlist}
\usepackage{tabularx}
\usepackage{tikz}
\usetikzlibrary{positioning}
\usepackage{xspace}
\usepackage{url}
\usepackage{titlesec}
\usepackage{subcaption}
\usepackage{tabularx}
\usepackage{array,graphicx}
\usepackage{booktabs}
\usepackage{pifont}
\usepackage{xspace}

 \newcommand{\naive}{na\"{\i}ve\xspace}

\renewcommand\footnotetextcopyrightpermission[1]{}
\begin{document}

\title{One Model to Rule them All:\\Towards Zero-Shot Learning for Databases}

\author{Benjamin Hilprecht}
\affiliation{
  \institution{Technical University of Darmstadt}
\country{Germany}}

\author{Carsten Binnig}
\affiliation{
  \institution{Technical University of Darmstadt}
\country{Germany}}

\begin{abstract}
In this paper, we present our vision of so called \emph{zero-shot learning for databases} which is a new learning approach for database components.
Zero-shot learning for databases is inspired by recent advances in transfer learning of models such as GPT-3 and can support a new database out-of-the box without the need to train a new model. Furthermore, it can easily be extended to few-shot learning by further retraining the model on the unseen database.
As a first concrete contribution in this paper, we show the feasibility of zero-shot learning for the task of physical cost estimation and present very promising initial results.
Moreover, as a second contribution we discuss the core challenges related to zero-shot learning for databases and present a roadmap to extend zero-shot learning towards many other tasks beyond cost estimation or even beyond classical database systems and workloads.
\end{abstract}

\maketitle

\section{Introduction}

\paragraph{Motivation.} Building computer systems often involves solving complex problems in all layers of those systems. 
To reduce complexity when building those computer systems and solve the problems in a tractable manner, these systems have heavily relied on heuristics or simplified analytical models in the past. 
Very recent work in the systems community, however, has outlined a broad scope where machine learning vastly outperforms these traditional heuristics.
This is also the case for databases, where existing DBMS components have been replaced with learned counterparts such as learned cost and cardinality estimation models \cite{kipf2019learned,10.14778/3384345.3384349,sun2019endtoend,10.14778/3421424.3421432,10.14778/3342263.3342646,DBLP:conf/cidr/HilprechtBBEHKR20} as well as learned query optimizers \cite{10.14778/3342263.3342644, marcus2020bao,Krishnan2018LearningTO,10.1145/3211954.3211957} or even learned indexes \cite{kraska2018thecase,10.1145/3318464.3389711,10.1145/3299869.3319860,10.14778/3425879.3425880} and learned query scheduling strategies \cite{DBLP:conf/sigmod/ShengTZP19,10.1145/3341302.3342080}.

The predominant approach that has been used in the past for learned database components is workload-driven learning.
The idea of workload-driven learning is to capture the behavior of a DBMS component by running a representative set of queries over a given database and then use the observations to train the underlying model.
For example, for learned cardinality estimation models such as \cite{kipf2019learned,sun2019endtoend} a set of queries must be executed to collect query plans and their cardinalities, which serve as training data for learning a model that can be used to estimate the cardinalities for new queries.
The very same procedure is applied if workload-driven learning is used for the other DBMS components such as learned physical design advisors (e.g., an advisor for index selection) \cite{10.1145/3299869.3300085, DBLP:journals/corr/abs-1903-01363} or other components. 

However, a major obstacle to these workload-driven approaches is the collection of training data.
For example, in \cite{kipf2019learned,sun2019endtoend} it was shown that thousands of query plans and their true cardinalities are needed for training the model to achieve a high accuracy. 
Running such a set of training queries on potentially very large databases to collect the training data can take hours or even days while the actual training of the underlying models often only takes a few minutes.
And unfortunately, the training data collection needs to be repeated for every new database that needs to be supported. 

To reduce the high cost of training data collection, reinforcement learning (RL) has been used to execute training queries \cite{10.1145/3299869.3300085, DBLP:journals/corr/abs-1903-01363, 10.1145/3318464.3389704, 10.14778/3199517.3199521} in a more targeted manner (i.e., letting the RL agent decide which queries to execute next). 
However, even with reinforcement learning still a large amount of training queries needs to be executed for learning a model.
Moreover, training the model is not a one-time effort since similar to workload-driven approaches the learning procedure needs to be repeated for every new database at hand.

A different direction that has thus been proposed to avoid the expensive training data collection by running queries on a new database are so called data-driven approaches \cite{10.14778/3384345.3384349,10.14778/3421424.3421432,Yang:2019:DUC:3368289.3374411} that learn a model purely from the underlying data.
A prime example where data-driven learning is a perfect fit is cardinality estimation.
However, data-driven learning is no silver bullet either since for some DBMS components the information about the runtime behavior of queries is required.
One such example is learned physical cost estimation where the runtime behavior of queries needs to be captured by a model to make predictions. 
A similar observation holds for many other database tasks such as physical design tuning or knob tuning where the effects of a certain decision on the runtime of a workload need to be learned.

\vspace{-1ex}\paragraph{Vision and Contributions.} In this paper, we thus present our vision of so called \emph{zero-shot learning for databases} which is a new learning approach for database components that can support a broad set of tasks on the one hand but does not require to collect training data for supporting a new database on the other hand.
In that regard, zero-shot learning for databases combines the benefits of data-driven learning and workload-driven learning.
The general idea behind zero-shot learning for databases is motivated by recent advances in transfer learning of models.
Similar to other approaches such as GPT-3 \cite{NEURIPS2020_1457c0d6} which enables zero-shot learning for NLP, a zero-shot model for databases is trained on a wide collection of different databases and workloads and can thus generalize to a completely new database and workload without the need to be trained particularly on that database.

As a core contribution in this paper, we discuss how such an approach of zero-shot learning for databases could work and we also show the feasibility of this approach for the task of physical cost estimation.
In our initial results for physical cost estimation, we show that zero-shot models can significantly outperform workload-driven approaches even when providing workload-driven models with a large number of training queries for a particular database at hand whereas zero-shot models can support them out-of-the-box.
Moreover, we believe that the real power of zero-shot learning stems from the fact that it is a general principle that can be used for various learned database tasks.
For example, we already have initial promising results suggesting that zero-shot learning can not only be used for physical cost estimation on a new database but also for physical design tuning and, in particular, index selection on a database the model has not seen before. 

Finally, an important aspect of zero-shot models is that zero-shot learning can easily be extended to \textit{few-shot learning}.
Hence, instead of using the zero-shot model out-of-the box (which already can provide good performance), one can fine-tune the model with only a few training queries on an unseen database or task.
Compared to workload-driven learning, few-shot learning will require way fewer training queries for adaptation since the general system behavior is already internalized by the zero-shot model.

\vspace{-1ex}\paragraph{Outline.}
The remainder of this paper is structured as follows:
Section~\ref{sec:overview} first gives an overview of zero-shot learning for databases and discusses the core challenges related to zero-shot learning.
Section~\ref{sec:cost} then discusses the case study of using our approach for learning a zero-shot physical cost model. Moreover, in this section we also present our initial experimental results.
Afterwards, Section~\ref{sec:beyond} discusses a research roadmap for zero-shot learning for databases beyond cost estimation.
Finally, we conclude with a peak into the future in Section~\ref{sec:concl}. \section{Zero-Shot Learning for Databases}
\label{sec:overview}

In this section, we first give a brief overview of how zero-shot learning for databases works in general and then discuss the core challenges related to enable this approach in an efficient manner.

\subsection{Overview of the Approach}

\begin{figure}
\centering
\includegraphics[width=\columnwidth]{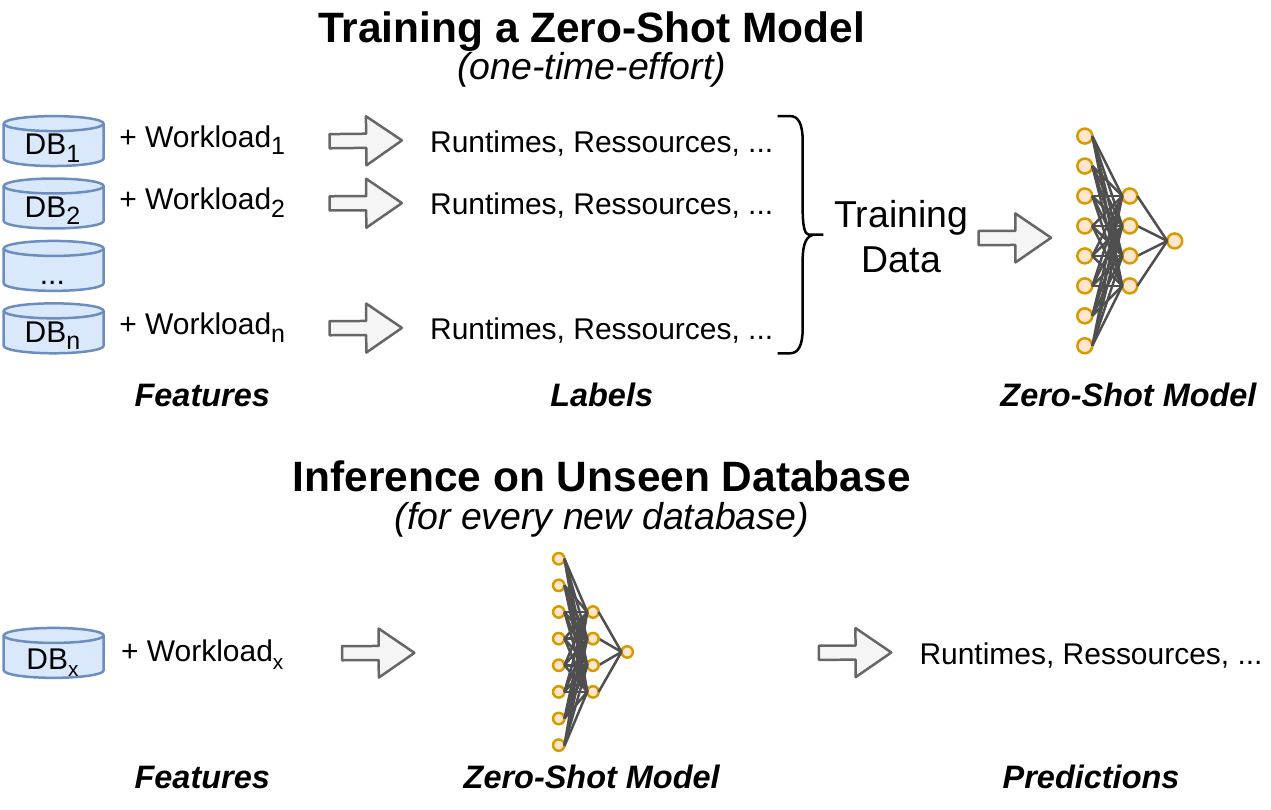}
\vspace{-4.5ex}
\caption{Overview of zero-shot learning for databases. In line with other zero-shot approaches such as GPT-3 which enables zero-shot learning for NLP, a zero-shot model for databases can generalize to a completely new database and workload without the need to be trained on that particular database.}
\vspace{-3.5ex}
\label{fig:overview}
\end{figure}

Figure \ref{fig:overview} shows the high-level idea that is behind our vision of zero-shot learning for databases.
During the learning phase, similar to workload-driven learning, for zero-shot learning we have to execute a representative workload and collect training data.

The main difference to workload-driven learning though, which makes our approach attractive, is that zero-shot models \emph{generalize to unseen databases out-of-the-box}. 
To allow a zero-shot model to make predictions about unseen databases without the need to retrain the model for this particular database, we provide a new method of representing queries as we discuss below (cf. Key Challenges).
This transferable representation is at the core of learning zero-shot models in a generalizable way and thus enables them to make predictions for queries on a new database (e.g., for physical cost estimation) that the model has never seen before.

Moreover, for being able to generalize to new databases, a zero-shot model is trained on different databases.
While this might seem to cause high upfront costs before a zero-shot model can be used, it is important to note that the \emph{training data collection is a one-time-effort} which is very different from workload-driven learning that needs to collect training data for every new database a model should support.
Moreover, cloud database providers such as AWS, Microsoft, or Google, typically already have significant amounts of such information available since they keep logs of their customer workloads and could thus apply zero-shot learning right away without the need to collect training data in the first place.

Finally, a last important aspect is that zero-shot learning is not only generalizable across databases but is a new learning approach that can be applied to a \emph{variety of database tasks} that range from physical cost estimation, design tuning or knob tuning to query optimization and scheduling as we discuss in Section \ref{sec:cost} and Section \ref{sec:beyond}.
To enable zero-shot models to generalize to different tasks though, the models need to be capable of capturing not only information about query plans and their runtimes but also information about other aspects (e.g., how indexes or changes in the database configuration influence the query runtime) as we discuss later.

\subsection{Key Challenges}

Enabling zero-shot learning for databases comes with various research challenges.
In the following, we discuss the key challenges that we think are at the core to make zero-shot learning for databases efficient and accurate. 

\vspace{-1ex}\paragraph{Transferable Representation of Database and Queries.}
State-of-the-art workload-driven models \cite{kipf2019learned, sun2019endtoend} can only leverage training data from a single database and thus they cannot simply be trained on a variety of databases to obtain zero-shot models. The reason is that the query representation is not \textit{transferable} to an unseen database. For instance, attributes names (e.g., those used in filter predicates) are typically encoded using a one-hot encoding assigning each column present in the database a specific position in a feature vector. Hence, the column \texttt{production\_year} of the IMDB dataset might be encoded using the vector $(0,1,0)$ (assuming that there are only three columns in total). If the same model is now used to predict query costs for the SSB dataset, the second column in the database might be \texttt{region}, which has very different semantics (i.e., very different data distributions or even a different data type). As such, cost models based on non-transferable representations will produce estimates that are most likely way off. In fact, such non-transferable feature encodings are used in various places of query representations such as table names as part of plan operators or literals in filter predicates. 
Hence, for zero-shot models we require a novel representation of queries that is transferable across databases while still being expressive enough to enable accurate estimations. 

\begin{figure}
	\centering
	\includegraphics[width=\columnwidth]{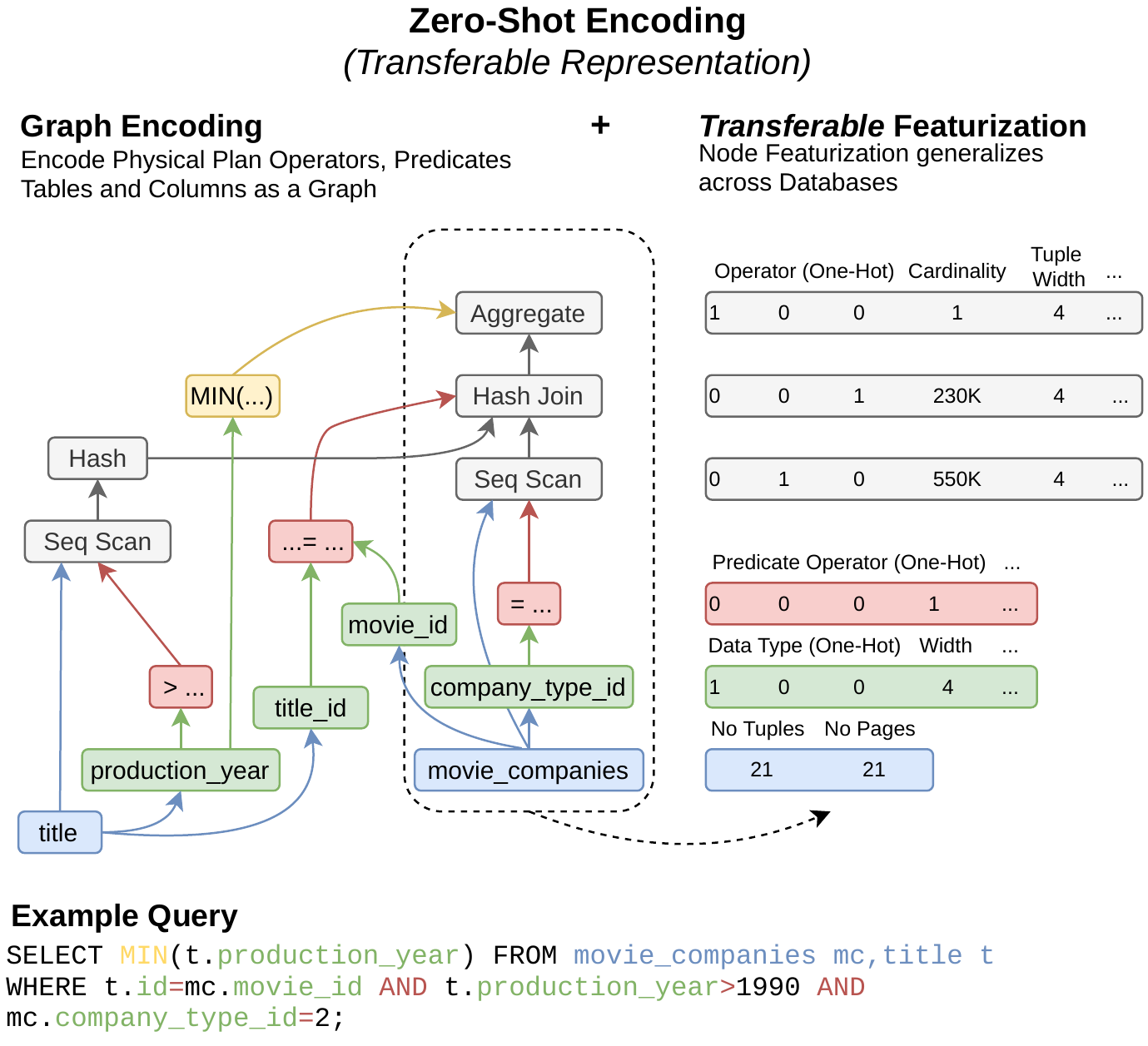}
	\vspace{-4.5ex}
	\caption{Graph encodings with transferable features as a zero-shot encoding. The query is represented as a graph with different node types (for plan operators, predicates, tables, columns etc.) and nodes are annotated with \textit{transferable} features. The representation allows the model to generalize across databases since features remain consistent.}
	\vspace{-3.5ex}
	\label{fig:encoding}
\end{figure}

We will now introduce \textit{query graph encodings with transferable features} (cf. Figure~\ref{fig:encoding}) which generalize across databases and have the potential to be applied in various zero-shot learned database components. While graph-based encodings have already been used to represent query plan operators and predicates \cite{sun2019endtoend}, we in contrast encode the entire query as a graph (including tables, columns etc.) and use transferable features per node allowing models using this representation to generalize across databases. For instance, an involved column \texttt{production\_year} would now be represented using a graph node with transferable features (e.g., the data type). If the same model is now deployed for a different database such as SSB, the corresponding columns would be represented using different graph nodes with their corresponding data characteristics and thus the representation is not inconsistent for different databases (in contrast to one-hot encodings of columns).

While the previously presented representation already allows to represent queries in a very expressive way, extensions might be necessary depending on the specific task that should be solved. When designing a representation for a specific task, it is important that the representation (i) captures all important aspects such that the models are able to solve the task at hand and (ii) that the representation is consistent for different databases (i.e., transferable).

\vspace{-1ex}\paragraph{Training Data Collection and Robustness.} 
A second key challenge to enable zero-shot learning is clearly the training data collection for learning a zero-shot cost model.
Here an important question is how many and which databases and workloads a zero-shot model needs to observe during training to make robust decisions on unseen databases.
As we show in our initial experiments for zero-shot cost estimation in Section~\ref{sec:cost}, for example, already a relatively small number of databases along with the respective workload information (e.g., the featurized query plans and runtimes) is sufficient to generalize robustly and even outperform existing workload-driven approaches.
However, clearly we need to develop more theoretical foundations to help guide us as to whether or not a zero-shot model has seen sufficiently many databases and queries.

One way to address this could be to evaluate the model on test databases that have not been used during training similar to the common practice in machine learning to evaluate a model on a holdout set. While this could provide an initial model validation, clearly more theoretical foundations are needed depending on the concrete task.  
A related question in this respect is how to create the training databases and workloads if they are not yet available (as for cloud providers).
An interesting direction here is to use a synthetic approach and generate databases / workloads with different characteristics.

\vspace{-1.5ex}\paragraph{Separation of Concerns.} 
Finally, a last important aspect of zero-shot models is to decide what should be learned by the model to fulfill its core promise and when to separate concerns.
For example, workload-driven approaches often prefer end-to-end learning, i.e., to make predictions for a query plan (e.g., the runtime), they internalize both the data characteristics (e.g., the data size and distributions) as well as the system characteristics (e.g., the runtime complexity of database operators) in one model. 

However, since the data characteristics can be entirely different for a new database, such an end-to-end approach will not work for zero-shot learning.
Hence, we suggest that data characteristics for zero-shot learning should be captured by separate data-driven models (such as \cite{10.14778/3384345.3384349,10.14778/3421424.3421432}).
For example, a feature that can be captured by a data-driven model are the input- and output-cardinalities of operators in a query plan. 
That way, when using cardinalities as input features for the zero-shot models, these models learn to predict the runtime behavior of operators based on input/output sizes that can be derived for any database which again enables a transferable representation of queries that does not depend on the concrete data distribution of a single database.

One could now argue that this violates the core promise of zero-shot learning since data-driven models need to be learned for each new database.
However, data-driven models can be derived from a database without running any training query and typically a sample of the database is enough to train these models.
Moreover, for cardinality estimation we could even use simple non-learned estimators (e.g., histograms) as input for the zero-shot models.
As we show in our initial results in Section \ref{sec:cost}, even those simple estimators often provide sufficient evidence for our zero-shot cost models to produce accurate estimates.

To summarize, a key question in this context is to decide what a zero-shot model should learn and which aspects should be treated separately.
Clearly a guide for this question is to think about what is tied to a particular data distribution and which aspects hold in general which should then be included in the zero-shot model. Moreover, as we discuss later, for design tuning or query optimization another question is how to combine zero-shot models with optimization procedures or other learning approaches (e.g., value networks) to implement efficient search strategies.  \section{Case Study: Cost Estimation}
\label{sec:cost}

In this case study, we demonstrate how zero-shot learning can be used for physical cost estimation. 
We envision this to be a potential core building block for zero-shot models for many other DBMS tasks as we discuss in the next section.

\vspace{-1.5ex}\subsection{Zero-Shot Cost Estimation}

The main promise of zero-shot cost estimation is that a trained model can predict the query runtime on a new database out-of-the-box.
In the following, we give a brief overview of how we implemented our initial prototype for zero-shot cost estimation for Postgres and contrast it to recent workload-driven approaches for cost estimation using the example query plan.

At the core, we use a \textit{transferable} query representation as introduced in Figure~\ref{fig:encoding} and propose a new architecture to capture the graph structure. As depicted in the figure, each node in this graph represents a physical operator (as opposed to a logical operator) to capture the differences in runtime complexity during learning. In addition, we use nodes to represent involved tables, columns, aggregations and predicates whereas for each node we use transferable features. For instance, the \texttt{movie\_companies} table uses generalizable features such as the number of tuples and pages. Note that for state-of-the-art workload-driven learning, the table would instead be represented as a one-hot encoded vector which does not enable generalization across databases as discussed before.

Moreover, as mentioned before, for zero-shot models we want the model to learn the general runtime behavior of operators in a DBMS (i.e., system characteristics). 
Hence, the zero-shot model should learn system characteristics separate from data characteristics.
This is very different from workload-driven models which learn both aspects end-to-end in one model as mentioned before.
For instance, workload-driven models include the values involved in filter predicates (e.g., \texttt{1990}) in the featurization of a query and thus learn the selectivity of the filter operation implicitly.
In contrast, for zero-shot models we only encode the predicate structure (to represent the general computational complexity) and explicitly use estimated cardinalities from a data-driven model or the query optimizer as input.

Finally, a last important aspect is our proposed model architecture as well as the learning and inference procedure for a featurized plan. Here, we exploit that the graph encoding results in DAGs where the root node of the plan are also the root nodes of the DAGs.
The learning overall happens in three steps: we first encode the features of the individual nodes in a fixed-size hidden vector (the initial hidden states). The hidden states of the individual plan nodes are afterwards combined using a bottom-up message passing phase in the DAG. Finally, the hidden state of the root node is fed into a multilayer perceptron (MLP) which predicts the final runtime.
In particular in the message passing phase, the DAG is traversed bottom up. The hidden states of the children are summed up (similar to the DeepSets \cite{NIPS2017_f22e4747} architecture) and combined with the hidden state of the parent node using an MLP to update the hidden state. This process is repeated until the root node is reached and the information of the entire tree is combined. 
For inference to make a prediction for a query plan on a new database, a new DAG is constructed using the node-specific MLPs and the features are propagated through the tree up to the root node which then predicts the runtime for the new query.

\begin{figure*}
	\centering
	\includegraphics[width=\linewidth]{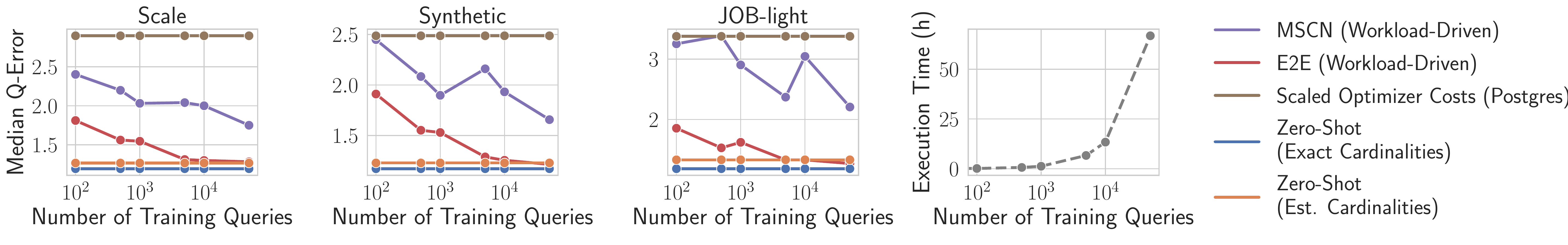}
	\vspace{-5.5ex}
	\caption{Estimation Errors of Workload-Driven Models for a varying Number of Training Queries compared with Zero-Shot Cost Models. The zero-shot models were trained using query executions on entirely different databases and thus do not require any queries on the IMDB database. In contrast, even the most accurate workload-driven model (E2E) requires approximately $10^4$ query executions on an unseen database for a comparable performance with zero-shot models which is roughly equivalent to $13$ hours of executed workload. Since zero-shot models do not require any additional queries it is significantly cheaper to deploy them for a new database.}
	\vspace{-3.5ex}
	\label{fig:exp_workload_size}
\end{figure*}

\vspace{-1.5ex}\subsection{Initial Evaluation}

\paragraph{Setup and Baselines} In an initial experiment, we trained a zero-shot cost estimation model on workloads we executed on a set of publicly available datasets that cover a range of databases with a different number of tables and database sizes. We then predicted the runtimes for a workload on an entirely unseen database to validate that zero-shot cost estimation can provide high accuracies.

As baselines, we used state-of-the-art workload-driven approaches for cost estimation. In particular, we used the end-to-end model of \cite{sun2019endtoend} called E2E and the MSCN architecture \cite{kipf2019learned} that was initially proposed for cardinality estimation. In addition, we use a simple linear model that obtains actual runtimes from the internal cost metric of the Postgres optimizer (called Scaled Optimizer Cost).

\vspace{-1ex}\paragraph{Training of Baselines and Zero-Shot Models} 
For the workload-driven models, we collected training data of different sizes ranging from a small training set size of $100$ queries up to very large training sets with $50,000$ queries (similar to \cite{sun2019endtoend}).
Note, that for every new database such training queries need to be executed and the runtimes need to be collected before a workload-driven model can be trained.
Moreover, if the database is updated and data characteristics change, the training data collection needs to be repeated.

For training the zero-shot model, we also need to collect training data. Overall, this gathering of the training data is clearly a significant effort. However, as mentioned before this is a one-time effort and the resulting model can be reused across different databases. 

To decide which number of training databases and workloads is sufficient, we evaluated the performance on a holdout test database as we added additional training databases. For each of the training databases we randomly generated $5,000$ training queries which we used as training data. After 19 databases, the performance stagnated and we can thus conclude that already a moderate number of training databases is sufficient for zero-shot models to generalize. 
In total the workload size for zero-shot learning was thus in a similar ballpark compared to the maximum size we used for training the workload-driven baselines.
Moreover, the queries for zero-shot learning and workload-driven learning were similar and covered up to five-way joins with up to five numerical and categorical predicates and up to three aggregates. 
However, different from a workload-driven model, the zero-shot model was not trained on the database it should make a prediction on.

\vspace*{-1.5ex}\paragraph{Initial Results} To evaluate the trained models, we used our zero-shot model as well as the other baselines to predict the runtimes of the commonly used \textit{scale}, \textit{synthetic} and \textit{job-light} benchmarks \cite{kipf2019learned} on the IMDB database. For zero-shot models, we show two versions: one that that uses the Postgres cardinalities as input as well as one version, which uses exact cardinalities (as an upper baseline to show how accurate zero-shot models can become). 

As a result, we report the commonly used Q-error which is the factor the predicted runtime deviates from the true runtime. The advantage of zero-shot learning over workload-driven approaches is that no queries on the test database are required for training. To demonstrate this tradeoff, we vary the number of training queries that can be used for the workload-driven baselines and compare the accuracy with zero-shot learning in Figure~\ref{fig:exp_workload_size}.
Overall, zero-shot learning can predict the runtimes very accurately. Since the IMDB dataset was never used for one of the training queries this shows, that zero-shot learning can generalize to unseen databases.
Interestingly, even the zero-shot model using only cardinality estimates of the Postgres optimizer is still very accurate. 

In contrast, the workload-driven E2E models \cite{sun2019endtoend} are less accurate than zero-shot models even for a large set of training queries on the IMDB database for the scale and synthetic benchmarks. For the simpler job-light queries that rarely contain range predicates, the E2E model is on par with the zero-shot model for the larger training sets. However, still the E2E models cannot match the performance of zero-shot learning with exact cardinalities. In addition, we note that the MSCN models are significantly less accurate since they use a much simpler featurization based on one-hot encodings (and not a tree-based featurization). Moreover, note that even though we repeated all measurements multiple times and report the median there are still some peaks in the reported Q-errors of MSCN due to a particularly high variance.

While our initial results are promising, in future we plan to conduct  more extensive experiments that show the robustness of zero-shot learning in several dimensions (e.g., more complex queries but also other databases).
 \section{Beyond Cost Estimation}
\label{sec:beyond}

In the following, we will discuss how zero-shot models can be extended beyond cost estimation.

\subsection{Physical Design and Knob Tuning} 
A first clear extension of zero-shot cost models as described in Section \ref{sec:cost} is to allow them to support a so called \emph{``What-If''} mode. This enables zero-cost models to predict the runtime of a query given a certain physical design or a database configuration (also called knob configuration).
For example, one could then ask the model how the runtime of a query changes if a certain index would exist or how the runtime changes if the buffer size would be increased.

Both these tasks --- physical design and knob tuning --- are classical problems of DBMSs that have been addressed in the past already by using optimization approaches \cite{10.5555/645926.671701,10.5555/645923.673646,10.1145/1066157.1066184,10.5555/1097871.1098137,10.1145/1989323.1989444}.
However, the main problem was that these approaches relied on inexact cost estimates coming from classical optimizer cost models that were extended to support a \emph{``What-If''} mode.
For that reason, recent approaches have suggested to use workload-driven learning --- in many cases reinforcement learning \cite{10.1145/3299869.3300085, DBLP:journals/corr/abs-1903-01363, 10.1145/3340531.3412106, 10.1145/3318464.3389704,Alabed2021HighDimensionalBO,10.1145/3437984.3458831,10.1145/3437984.3458830}.
While these approaches have been shown to be more accurate than the more classical approaches, they again need to first run training queries under different physical layouts or knob configurations for every new database.

Hence, to avoid these high-training costs for every new database one could use a zero-shot cost model in a \emph{``What-If''} mode.
To show the feasibility of this direction, we extended our zero-shot cost model to support also decisions towards index tuning.
At the core, the zero-shot model for index tuning should be able to support predictions of the runtime as if a certain index would exist in a database.
For training a zero-shot cost model that can answer such questions, we again used the 19 databases as training data that we also used in Section \ref{sec:cost}.
However, we additionally created a random but fixed set of indexes per database before running the training queries. The zero-shot cost model could recognize for which training query an index was used since the physical operators in a query plan change (e.g., an index scan instead of a table scan was used). 
The zero-shot model could thus learn how the runtime changes for query plans, which use an index scan compared to query plans which do not use an index scan.

For showing that the learned model could estimate the runtime of queries correctly given a certain index, we again use the IMDB database for the evaluation that the zero-shot model had not seen before.
For testing, we asked the model to estimate the runtime of queries if an index would exist again for randomly selected attributes of queries. The estimation errors for zero-shot learning for this workload are given in Table~\ref{fig:q_errors} (last line).
As we can see, the estimations are still very accurate but clearly the maximum Q-error increases compared to the results for zero-shot cost models in Section \ref{sec:cost} before (upper lines).

To further improve the quality of zero-shot cost models for index tuning one might need to think about a more sophisticated workload sampling or provide additional characteristics for indexes (e.g., expected index height) as input features to a zero-shot model that could be derived with additional data-driven models.
Furthermore, we think that we could use zero-shot models to build other design advisors (e.g., for materialized views) or support zero-shot knob tuning to select an optimal database configuration for a given workload without having seen the database for training. 
Finally, for knob tuning one could also think about using zero-shot models to only guide the search initially to a good start configuration (i.e., to narrow down the search space) and then use online approaches for fine-tuning the knobs since knobs can be changed easily compared to the high cost of changing a physical layout.

\begin{table}
	\scriptsize
	\centering
	\begin{tabular}{lllllll}\toprule
		& \multicolumn{3}{l}{Zero-Shot (Exact Card.)} & \multicolumn{3}{l}{Zero-Shot (Estimated Card.)} \\
		Workload & median & 95th & max & median & 95th & max \\
		\midrule
		Scale & 1.19 & 1.93 & 3.93 & 1.26 & 2.46 & 4.70 \\
		Synthetic & 1.17 & 1.90 & 4.40 & 1.21 & 2.17 & 6.88 \\
		JOB-light & 1.18 & 1.85 & 2.47 & 1.33 & 2.56 & 4.00 \\
		Index & 1.21 & 2.51 & 10.73 & 1.33 & 3.59 & 24.62 \\
		\bottomrule
	\end{tabular}
\caption{Estimation errors (Q-errors) of zero-shot models for index tuning (last line) compared to zero-shot cost models without \emph{What-if} support (upper lines).}
	\vspace*{-7.5ex}
	\label{fig:q_errors}
\end{table}

\subsection{Query Optimization} 
Another direction for zero-shot learning are end-to-end learned optimizers and not just the learning of cost models.
Recently, it was also proposed to replace query optimizers that typically rely on heuristics (i.e., simple cost models) and manual engineering by machine learning models \cite{10.14778/3342263.3342644, marcus2020bao,Krishnan2018LearningTO,10.1145/3211954.3211957}. While initial results are promising and even commercial optimizers can be outperformed by learned ones, current approaches are also dominated by reinforcement leaning or workload-driven learning in general. Again, all these approaches are database-dependent and cannot generalize to unseen databases.
Moreover, for learning an optimizer a huge number of queries has to be executed to learn what is a good plan for a given query. We envision that this overhead for unseen databases can be eliminated completely using zero-shot learning. 

An initial \naive{} approach for this could be to use the devised zero-shot cost estimation model to evaluate candidate plans and thus better guide the query optimizer to plans with low costs. For instance, zero-shot cost estimation models could be used in conjunction with classical dynamic programming or even approaches like Bao \cite{marcus2020bao}. However, with more sophisticated approaches, we think that zero-shot learning could potentially replace classical heuristics like dynamic programming entirely by devising zero-shot value networks to learn search strategies for query optimization. 
Value networks \cite{DBLP:conf/nips/TamarLAWT16} have shown to learn policies that involve planning-based reasoning.
This way, zero-shot query optimizers could come up with plans that classical optimizers would not have considered while avoiding the burden to run thousands of queries to train the learned optimizer for every new database.

\vspace{-1.5ex}\subsection{Discussion} 

In addition to design advisors, knob tuning, or database optimizers there are many more DBMS components that could benefit from zero-shot learning. 
For example, zero-shot cost models could be used to predict not only the runtime but also other aspects such as resource consumption and thus be used also for runtime decisions (e.g., query scheduling).
Moreover, by extending the features of the \emph{``What-If''} mode, we could also support hardware aspects and predict the runtime of queries on an unseen hardware, e.g., to select an optimal cloud instance for a given workload.

Another interesting question is how zero-shot learning should be integrated into the overall DBMS architecture.
Here we envision a route where \emph{zero-shot cost models} as presented in Section \ref{sec:cost} form a ``kind-of'' \emph{central brain} in a DBMS that can be leveraged by various DBMS components that complement such a central component with more targeted models. 
These additional models could for example be zero-shot models that focus on learning particular search strategies or specific data-driven models to capture interesting data characteristics as we discussed before.

Finally, as mentioned before it can be beneficial to fine-tune a zero-shot model also on the unseen database. The resulting few-shot models leverage the observed workload on the database similar to workload-driven models and thus likely offer more accurate predictions. However, the main difference to workload-driven models is that our approach also offers accurate out-of-the-box predictions for unseen databases by using zero-shot models and also requires fewer queries for adaptation on an unseen database since the general system behavior is already internalized by the zero-shot model. As such, it is significantly more efficient to fine-tune a model for unseen databases than to train one from scratch every time as it is necessary for workload-driven models.
 \section{Looking into the Future}
\label{sec:concl}

In this paper, we have shown a new approach for learned database components that can support new databases without running any training query on that database. Moreover, zero-shot models can be fine-tuned on the unseen database for more accurate predictions resulting in few-shot models.
While we have focused on single-node databases and classical database workloads in the first place, we believe that zero-shot models can be applied more broadly.
One direction are distributed DBMSs where zero-shot models can be extended to support tasks such as to optimize a distributed data layout.
Another direction is to extend zero-shot models for other types of data-intensive workloads (e.g., data streaming). 

Moreover, when thinking more broadly, zero-shot models seem to also be an attractive model for any system builder and can be also used at various levels of granularity to predict the performance of individual components (e.g., very fine-grained on the data structure and algorithm level) or very coarse-grained (at the system level).
For example, when being used for data structures and algorithms, zero-shot models would be an efficient vehicle for self-designing data structures \cite{DBLP:journals/debu/IdreosZCQWHKDGK19}.
To conclude, we think that zero-shot learning opens up many avenues of research since it provides not only a more \emph{sustainable} way to build learnable system components but it also seems to be a \emph{general paradigm} that can be applied more broadly and at different levels. 

\section{Acknowledgments}

We thank the reviewers for their feedback and comments.
This research and development project is funded by the German Federal Ministry of
Education and Research (BMBF) within the “The Future of Value Creation – Research on Production, Services and Work” program and managed by the Project
Management Agency Karlsruhe (PTKA). The author is responsible for the content of this
publication. In addition, the research was partly funded by the Hochtief project \emph{AICO} (AI in Construction), the HMWK cluster project \emph{3AI} (The Third Wave of AI), as well as the DFG Collaborative Research Center 1053 (MAKI). Finally, we want to thank the Amazon Redshift team for valuable discussions.
 
\balance{}

\begin{scriptsize}
\bibliographystyle{abbrv}
\bibliography{bib}
\end{scriptsize}

\end{document}